\documentclass[prl,twocolumn,showpacs,preprintnumbers,amsmath,amssymb,floatfix]{revtex4-2}

\usepackage{graphicx}
\usepackage{dcolumn}
\usepackage{bm}
\usepackage{color}
\usepackage[utf8]{inputenc}
\usepackage{hyperref}

\newcommand{\affA}{Department de F\'isica, Universitat Polit\`ecnica de Catalunya, Campus Nord B4-B5, E-08034, Barcelona, Spain}
\newcommand{\affB}{Institut f\"ur Theoretische Physik, Leibniz Universit\"at Hannover, Appelstr. 2, 30167 Hannover, Germany}
\newcommand{\affC}{Max Planck Institute for Quantum Optics, 85748 Garching, Germany}
\newcommand{\affD}{Munich Center for Quantum Science and Technology (MCQST), Schellingstra\ss e 4, D-80799 M\"unchen, Germany}
\newcommand{\affE}{Forschungszentrum J\"{u}lich Gmbh, Institute of Quantum Control (PGI-8), D-52425 J\"{u}lich, Germany}
\newcommand{\affF}{Faculty of Physics, University of Warsaw, Pasterua 5, PL-02093 Warsaw, Poland}
\newcommand{\affG}{Zentrum f\"ur Optische Quantentechnologien, Fachbereich Physik, and The Hamburg Centre for Ultrafast Imaging, Universit\"at Hamburg, Luruper Chaussee 149, D-22761 Hamburg, Germany}

\begin{document}

\title{Ionic polaron in a Bose-Einstein condensate}
\author{Grigory E. Astrakharchik$^{1*}$, Luis A. Pe\~na Ardila$^2$, Richard Schmidt$^{3,4}$, Krzysztof Jachymski$^{5,6}$, Antonio Negretti$^7$}
\affiliation{$^1$\affA}
\affiliation{$^2$\affB}
\affiliation{$^3$\affC}
\affiliation{$^4$\affD}
\affiliation{$^5$\affE}
\affiliation{$^6$\affF}
\affiliation{$^7$\affG}
\affiliation{$^*$Corresponding author,grigori.astrakharchik@upc.edu}
\date{\today}

\begin{abstract}
{\bf Abstract}
The presence of strong interactions in a many-body quantum system can lead to a variety of exotic effects. Here we show that even in a comparatively simple setup consisting of a charged impurity in a weakly interacting bosonic medium the competition of length scales gives rise to a highly correlated mesoscopic state. Using quantum Monte Carlo simulations, we unravel its vastly different polaronic properties compared to neutral quantum impurities. Moreover, we identify a transition between the regime amenable to conventional perturbative treatment in the limit of weak atom-ion interactions and a many-body bound state with vanishing quasi-particle residue composed of hundreds of atoms. In order to analyze the structure of the corresponding states we examine the atom-ion and atom-atom correlation functions which both show nontrivial properties. Our findings are directly relevant to experiments using hybrid atom-ion setups that have recently attained the ultracold regime.
\end{abstract}

\maketitle


{\bf Introduction}  

An impurity immersed in a many-body quantum system constitutes a fundamental building block in condensed-matter physics, particularly with regards to transport properties of materials~\cite{Mahan2000,Alexandrov2003}. In order to investigate this paradigmatic problem, ultracold atoms are especially suited as they allow for experimental control of multiple parameters such as the shape of the confinement and the form of interparticle interactions~\cite{Bloch2008}. Over the last decade atomic systems have proven to be capable to observe the formation of quasi-particles such as the celebrated polaron in bosonic~\cite{Hu2016,Jorgensen2016,Yan190} and fermionic~\cite{SchirotzekPRL2009,ZhangPRL2012,KoschorreckNature12,KohstallNATURE2012,Cetina2016,ScazzaPRL17} quantum gases as well as in a Rydberg system~\cite{Camargo2018}, with the possibility of exploring impurity physics also in the presence of dipolar interactions~\cite{Trautmann2018}. 

Within the plethora of compound atomic quantum systems available, atom-ion systems provide a unique arena for investigating many-body quantum physics in the strongly interacting regime. Indeed, the interaction between the charge and the induced dipole of the neutral particle 
results in the asymptotic form 
\begin{align}
\label{eq:Vai}
V_{\mathrm{ai}}(\mathbf{r}) \stackrel{r\to\infty}{\longrightarrow} -\frac{C_4}{r^4}\, .
\end{align}
Importantly, this polarization potential has a characteristic length scale that is about an order of magnitude larger than in the case of van der Waals interactions typical for neutral atoms and it can become comparable to the mean interparticle distance. Moreover, the characteristic interaction energy is typically in the microkelvin range and thus comparable to experimentally achievable collision energies~\cite{TomzaRMP19}. Hence, although the polarization potential has in principle short-ranged nature one cannot replace it with a pseudopotential due to the lack of separation of length and energy scales between the two-body and the many-body systems. While this lack of scale separation gives rise to a striking competition of few- and many-body processes, it poses a theoretical challenge due to the necessity to account for details of the potential that severely inhibits the possibility of using analytical methods to describe the properties of an ionic impurity such as its effective mass.

Recently, it has been demonstrated that for certain atom-ion combinations the ultracold regime is within experimental reach~\cite{Kleinbach2018,Feldker2019,SchmidtPRL20}. Exploiting the charge of the impurity, one appealing possibility is to study transport properties by dragging the ion by means of electric fields and detect it with high spatial and temporal resolution~\cite{dieterle2020transport}. The long range of the atom-ion potential, on the other hand, allows to investigate the formation of mesoscopic bound states~\cite{CotePRL02,GaoPRL2010,SchurerPRL17} that would have dramatic impact on the ion transport dynamics. 

Studies of the mobility of an ion moving in a bosonic medium date back to the early sixties with the aim of explaining the small ion mobility in liquid helium~\cite{WilliamsCJP1957,MeyerPR1958,CareriNC1959,AtkinsPR1959,GROSS1962234}. Later, mean-field approaches were used to predict the formation of mesoscopic molecular ions~\cite{CotePRL02}, to estimate the number of excess atoms around an ion in a homogeneous Bose-Einstein condensate (BEC)~\cite{MassignanPRA05}, and path integral methods to determine the ion polaron properties in the strong-coupling regime~\cite{CasteelsJLTP11}. While such approaches allow one to obtain a qualitative picture of the underlying phenomenology, the interplay between the long-ranged potential and strong interactions not only leads to substantial shifts to the relevant observables such as the energy of the system, but also has drastic consequences for the structural properties of the ground state. In line with the recent experiments with ion-atom mixtures~\cite{SchmidPRL10,ZipkesNature10,HallPRL11,GoodmanPRA12,RaviNatCommun12,MeirPRL16,Wessels2017,Meir2018,Kleinbach2018,Engel2018}, in this letter we study the many-body ground state properties of an ion immersed in a three-dimensional (3D) bosonic gas. To this end, we employ quantum Monte Carlo techniques that have been successful in the context of the 3D Bose polaron~\cite{ArdilaPRA2015,ArdilaPRA2016,ArdilaPRA2019}, bipolarons~\cite{ArdilaPRL2018}, as well as in two-dimensional (2D)~\cite{Ardila2020} and one-dimensional (1D) polarons~\cite{AstrakharchikBrouzos2013,Grusdt_2017,Parisi2017}. The method allows us to fully take into account the quantum many-body correlations that turn out to be important for predicting how the competition between the few- and many-body length scales gives rise to a striking impurity physics that is governed by a transition from a polaron to a many-body polaron bound state. The resulting states cannot be captured by conventional tools such as the Fr\"{o}lich model or Bogolyubov theory. The system properties depend not only on the scattering length and effective range of the two-body potential, but rely on the presence of the long-range tail of the interaction, indicating the failure of the pseudopotential approximation.

\begin{figure}
\centering
\includegraphics[width=\columnwidth]{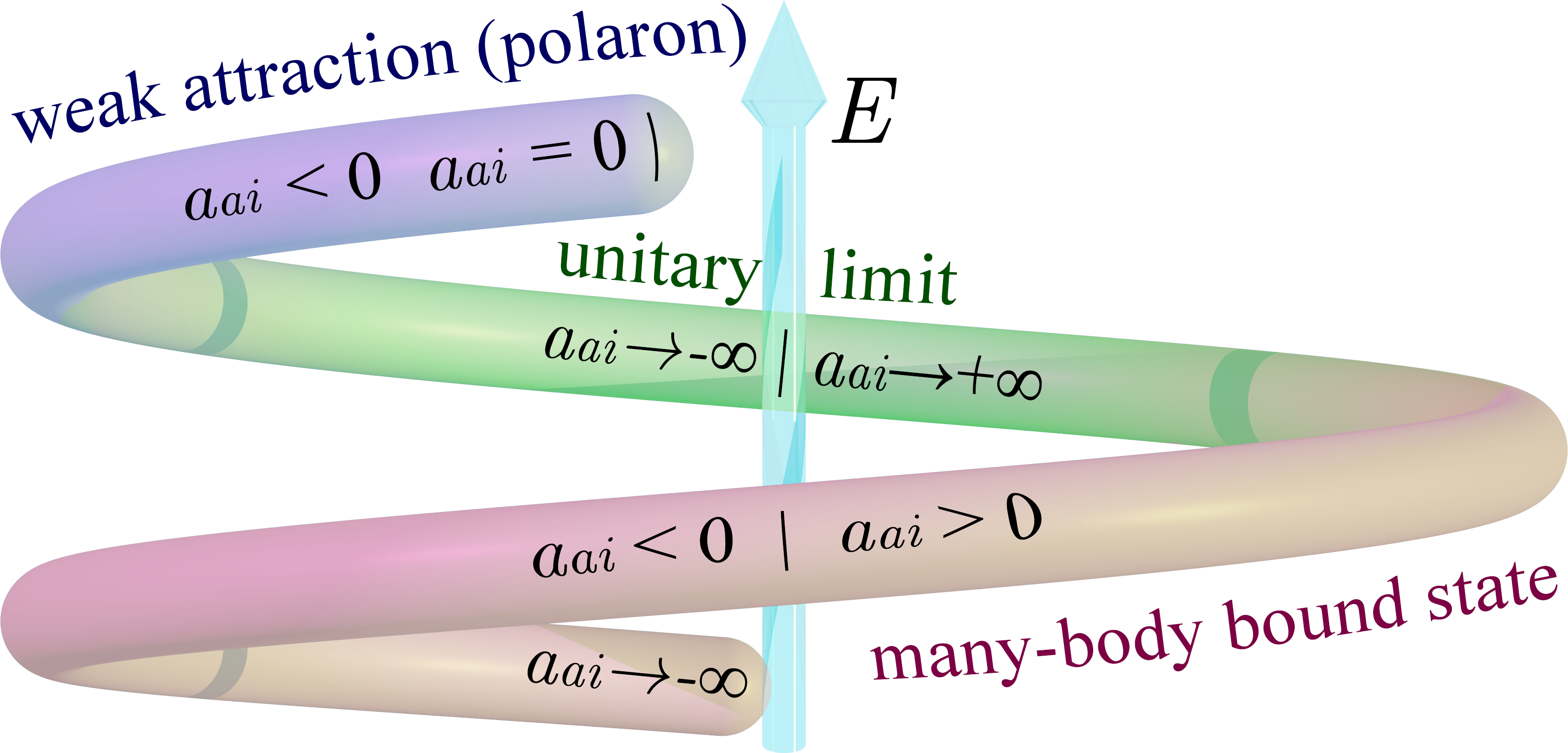}
\caption{\label{fig:sketch}
{\bf Schematic phase diagram.}
A schematic phase diagram showing various regimes for a single ion immersed in a dilute Bose gas upon changing the atom-ion scattering length and the number of the two-body bound states.
The vertical axis indicates the change of the ion energy $E$ as the value of atom-ion $s$-wave scattering length $a_{ai}$ is varied. The spiral refers to a continuous change of the energy leading to realization of different physical regimes as the value of $a_{ai}$ is cyclically changed from minus to plus infinity. Departing from the non-interacting case $a_{ai}=0$, with zero ion energy, $E=0$, and by making the ion attraction stronger the following regimes are observed. A weak attraction leads to the polaronic regime where the ion can be described in terms of a quasiparticle. Such a description is no longer possible in the unitary regime marked by a diverging $s$-wave scattering length. 
The unitary point, $|a_{ai}|=\infty$, is the threshold value for the formation of an ion-atom two-body bound state. 
For stronger atom-ion attraction  many-body bound state is formed in which a large number of atoms is effectively trapped by the ion.
} 
\end{figure}

{\bf Results}

{\it System.} We consider an ion of mass $M$ immersed into a gas consisting of $N$ bosonic atoms of mass $m$ at average density $n=N/L^3$. For simplicity we focus here on the mass-balanced case (i.e. $M=m$). We consider periodic boundary conditions in a box of size $L$ chosen large compared to the healing length $\xi = (8\pi n a_{\mathrm{aa}})^{-1/2}$, where $a_{\mathrm{aa}}$ is the boson-boson $s$-wave scattering length.

The microscopic many-body Hamiltonian is given by
\begin{align}
\label{eq:H}
\hat H\!=\! 
-\frac{\hbar^2\nabla^2_{\mathbf{R}}}{2 M} \!-\!\!\sum_{n=1}^N \!\frac{\hbar^2\nabla^2_{\mathbf{r}_n}}{2 m} -V_{\mathrm{ai}}(\mathbf{r}_n\!-\!\mathbf{R})
\!+\!\sum_{n<j}^NV_{\mathrm{aa}}(\mathbf{r}_n\!-\!\mathbf{r}_j).\!
\end{align}
Hereafter, we denote the ion's characteristics such as position and mass with capital Latin letters, while for atom ones we use small Latin letters. Furthermore, bold symbols refer to three-dimensional vectors and cursive ones the respective norms.
The first two terms in Eq.~(\ref{eq:H}) represent the kinetic energy of the ion and of the atoms, respectively, and $V_{\mathrm{aa}}(\mathbf{r}_n-\mathbf{r}_j)$ is a repulsive short-range potential with coupling constant $g = 4\pi\hbar^2 a_{\mathrm{aa}}/m$. The potential $V_{\mathrm{ai}}$ describes the atom-ion interaction, for which it is essential to retain the long-range tail~(\ref{eq:Vai}).  It is further characterised by the length $R^\star = (2 m_{\mathrm{r}} C_4/\hbar^2)^{1/2}$ and energy scales $E^\star= \hbar^2 / [2m_{\mathrm{r}} (R^\star)^2]$, where $m_{{\mathrm{r}}} = m M/(m+M)$ is the reduced mass. For the $^{87}$Rb/$^{87}$Rb$^+$ system one has $R^\star\simeq 265.81$ nm and $E^\star\simeq k_{\mathrm{B}}\times 79$ nK ($k_{\mathrm{B}}$ is the Boltzmann constant). Importantly, the separation of length scales is lacking as for typical atom density $n = 10^{14}$cm$^{-3}$ the mean interparticle distance $n^{-1/3}\simeq 0.8\,R^\star$ is of the same order as the interaction range as well as the healing length ($\xi\simeq R^\star$).

At short range, the real interaction potential will deviate from the asymptotic formula~\eqref{eq:Vai}. Here, we model the short-range details by the regularization~\cite{KrychPRA15}:
\begin{align}
\label{eq:Vaireg}
V_{\mathrm{ai}}^{r}(\mathbf{r}) = -C_4\frac{r^2 - c^2}{r^2 + c^2} \frac{1}{(b^2 + r^2)^2}.
\end{align}
Here, the $b$ and $c$ parameters have units of length and control the properties of the potential such as the number of bound states and their energies as well as the scattering length, while the long-range effects of the tail~\eqref{eq:Vai} remain accounted for. Crucially, the properties of the system depend not only on the energies, but also on the number of the available two-body bound states, as it will be demonstrated below. In most of the calculations we tune the potential in such a way that it has only one two-body bound state. Under this assumption, there is a unique connection between the $b$, $c$ parameters and the $s$-wave scattering length $a_{\mathrm{ai}}$ of the resulting potential. 

The simulations are performed by using variational (VMC) and diffusion Monte Carlo (DMC) methods. The VMC method samples the square of the trial wavefunction that we choose in the Bijl-Jastrow~\cite{Bijl1940,JastrowPR1955} form
\begin{align}
\label{eq:Psi-Trial}
\Psi_{\mathrm{T}}(\mathbf{R};\mathbf{r}_1,\dots,\mathbf{r}_N) = 
\prod_{j<n}f_{\mathrm{B}}(\vert\mathbf{r}_n - \mathbf{r}_j\vert) 
\prod_{j<n}f_{\mathrm{I}}(\vert\mathbf{r}_n - \mathbf{R}\vert). 
\end{align}
Here, $f_{\mathrm{B}}$ and $f_{\mathrm{I}}$ account for two-particle intra- and inter- species correlations, respectively. These functions are constructed by matching the solution of the two-body scattering problem at short distances to an appropriate tail (see Methods), i.e. phononic decay in $f_{\mathrm{B}}$ and mean-field prediction for a heavy ion in $f_{\mathrm{I}}$. Both functions contain variational parameters that are optimized by minimizing the expectation value of the Hamiltonian~(\ref{eq:H}). VMC calculations provide the upper bound to the ground-state energy. In contrast the DMC approach aims to obtain the exact ground state energy of the system by solving the Schr\"odinger equation in imaginary time. 
We are interested in the regime of weak atom-atom interactions and fix the gas parameter to $n a_{\mathrm{aa}}^3 = 10^{-6}$ with $a_{\mathrm{aa}} = 0.02\,R^\star$. 


{\it Ground state energy.} 
In Fig.~\ref{fig:sketch} we illustrate the `phase diagram' of the system that is characterized by two distinct sets of ground states: many-body bound state (MBBS) and a polaronic one. 
A spiral is used to illustrate that the energy $E$ depends monotonously on the $s$-wave scattering length $a_{\mathrm{ai}}$, although different energies correspond to the same value of $a_{\mathrm{ai}}$ depending on the number of bound states.
While for $a_{\mathrm{ai}}>0$ the potential~(\ref{eq:Vaireg}) always has a bound state, for negative scattering lengths the atom-ion interaction can be tuned such that either a bound state is supported (left-bottom part of the helix, MBBS), or no bound state is present (left-upper part, polaronic). The `attractive' polaron is typically encountered in ultracold quantum gases with neutral atomic impurities~\cite{RathPRA13,ArdilaPRA2015,ArdilaPRA2016,Hu2016,Jorgensen2016}. On the other hand, in the MBBS regime many bosons can be bound to the ion with a large binding energy. Importantly, the spatial range of the atom-ion interaction plays a crucial role in the formation of the MBBS, while for neutral impurities physics can typically be well described by assuming an effective short-range interaction. 

\begin{figure}
\centering
\includegraphics[width=\columnwidth]{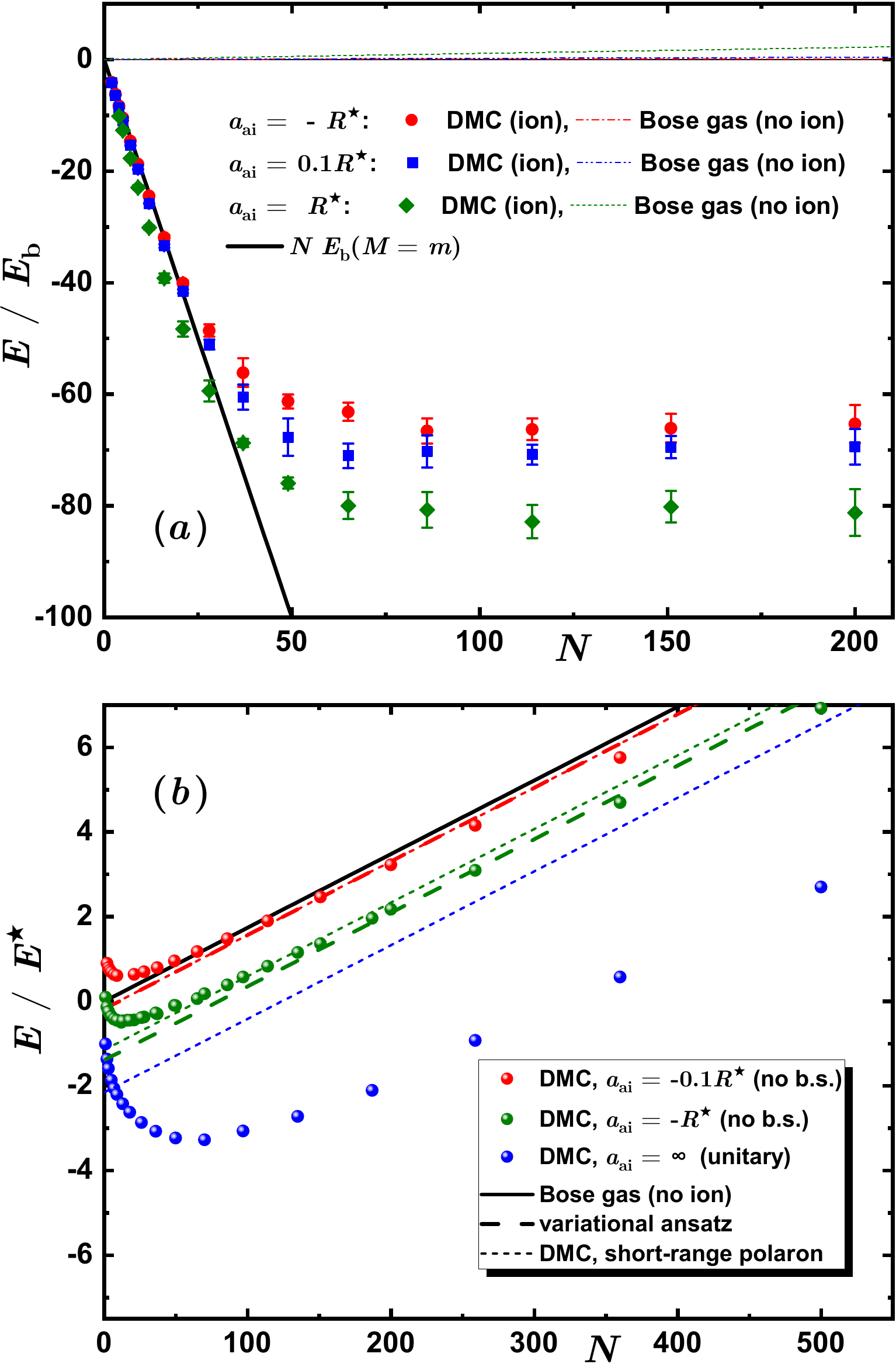}
\caption{\label{fig:energy} 
{\bf Ground-state energy of the many-body system.} 
(a) Total energy obtained as the expectation value of the Hamiltonian~(\ref{eq:H}) in units of the binding energy of the ion-atom molecule $E_{\mathrm{b}}$ as a function of the number of bosons $N$. 
The symbols represent the energy obtained in diffusion Monte Carlo (DMC) simulations, while the black solid line shows the energy~(\ref{eq:mu:pinned}). 
Error bars show the standard deviation of Monte Carlo simulations. 
(b) Total energy in units of $E^{\star}$. 
For the lines, the mean-field Gross-Pitaevskii energy is shifted by the polaron energy, i.e. $E(N) = \mu_{\mathrm{pol}} + g n N/2$;
solid line, no ion, $\mu_{\mathrm{pol}}=0$; 
long-dashed line, variational ansatz~\cite{ShchadilovaPRL16}, $\mu_{\mathrm{pol}}$ is given by Eq.~(\ref{eq:mu:variational});
short-dashed line, diffusion Monte Carlo calculation for short-ranged impurity-boson interactions from Ref.~\cite{ArdilaPRA2015}.
}
\end{figure}

\begin{figure}
\centering
\includegraphics[width=\columnwidth]{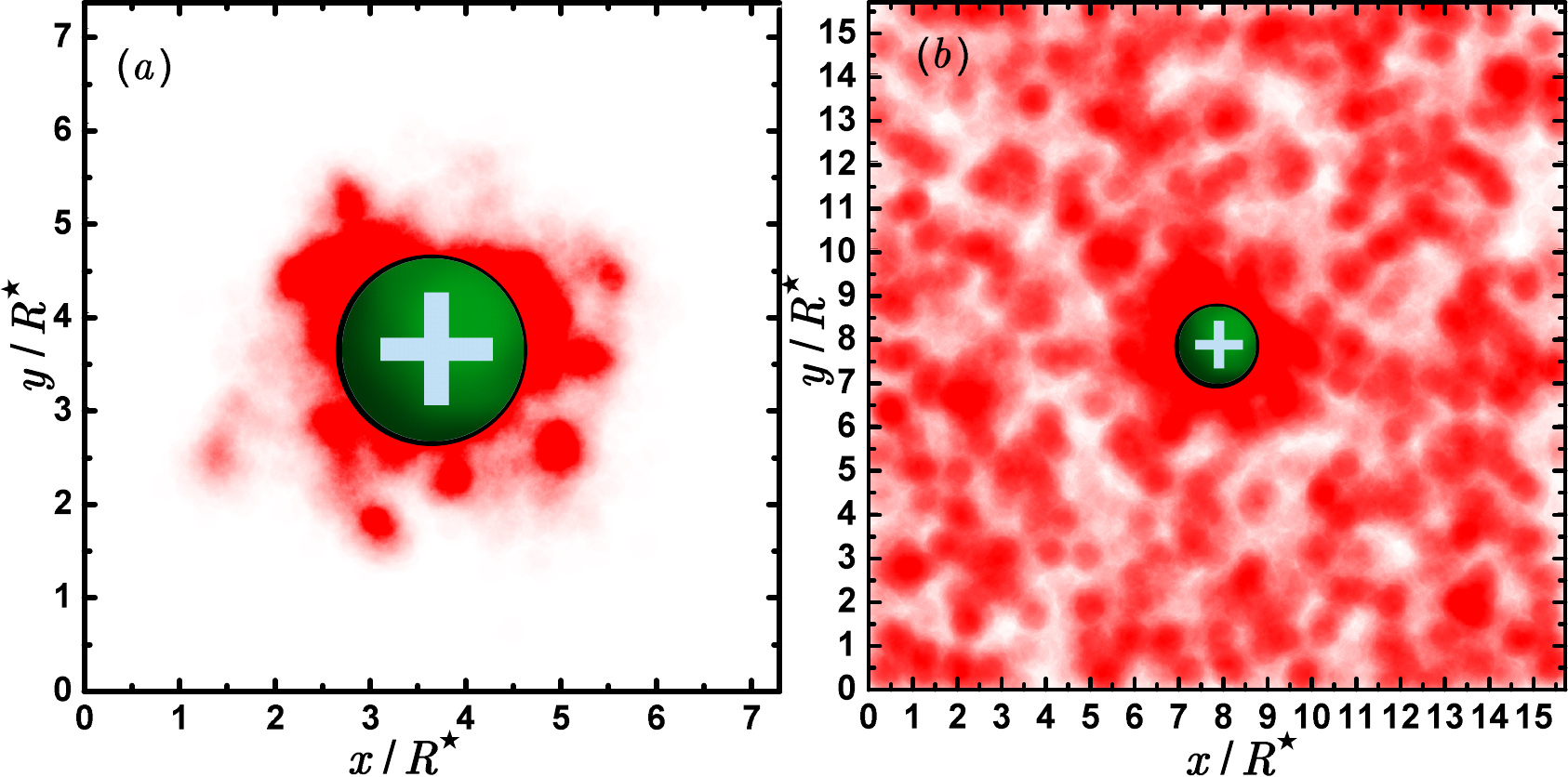}
\caption{\label{fig:snapshots} 
{\bf Snapshots of instantaneous particle positions in numerical simulations. }
Particle coordinates are represented for two characteristic numbers of bosons:
(a) $N=50$ particles (b) $N=500$ particles. 
Note that in the critical number of atoms which can be trapped by the ion, $N_{\mathrm{c}}=140$, is (a)~larger (b)~smaller than the number of particles $N$, correspondingly. 
Parameters of the interaction are chosen such that the value of the atom-ion $s$-wave scattering length equals to the characteristic length-scale $R^\star$ of the $C_4$ decay~(\ref{eq:Vai}), $a_{\mathrm{ai}} = R^\star$. The red areas denote instantaneous positions of the bosons in the box potential. Coordinates are shifted so that the ion appears in the center. 
}
\end{figure}

Figure~\ref{fig:energy} provides characteristic examples of the dependence of the system's total energy on the number of bosons in the MBBS (a) and polaronic (b) regime. In the MBBS case, we find that the absolute value of the energy grows almost linearly for a sufficiently small number of bosons. The dependence can be roughly approximated by the energy of $N$ non-interacting particles bound to the ion,
\begin{equation}
E(N) \simeq N E_{\mathrm{b}}(M=m)\;,
\label{eq:mu:pinned}
\end{equation}
as shown with a solid black line in Fig.~\ref{fig:energy} (a). We also have verified that the effective impurity mass approaches the total mass of the MBBS, $M^\star \approx Nm$. As the number of bosons is increased further, the energy starts to significantly deviate from the behavior~(\ref{eq:mu:pinned}) and it reaches a minimum at a critical number $N_{\mathrm{c}}$, which can be estimated from the extremum condition $\partial E/\partial N = 0$. The value $N_{\mathrm{c}}$ can be interpreted as the maximal number of bosons bound to the impurity, similarly to the analysis of the 1D case~\cite{SchurerPRL17}. We note that $N_{\mathrm{c}}$ could be defined in different ways, e.g. from the form of the atom-ion correlation function, $g_2^{\mathrm{ai}}(r)$, as $N_{\mathrm{c}} = n\int[g_2^{\mathrm{ai}}(r)-1+N_{\mathrm{c}}/N]4\pi r^2 dr$, or from the atom density far from the ion, i.e. $n (1-N_{\mathrm{c}}/N)$. For the chosen parameters we obtain $N_{\mathrm{c}}\simeq 140$ almost irrespective of the exact value of the atom-ion scattering length, contrarily to the mean-field prediction of Ref.~\cite{CotePRL02}. This indicates that while the scattering physics of quasi-free bosons is determined by the scattering length, it is the large range of the potential that determines the number of bound particles. Note that our result is significantly larger as compared to the many-body bound states for the case of a neutral impurity, for which Monte Carlo calculations predict only few atoms to be bound~\cite{ArdilaPRA2015}. At the same time $N_{\mathrm{c}}$ is much smaller than the number of bound atoms predicted by a mean-field-based estimate, suggesting that the effective gas parameter is significantly increased in the vicinity of the ion. For $N > N_{\mathrm{c}}$ the energy increases, meaning that no more bosons are able to bind to the ion and the excess atoms start to form an almost uniform gas. This view is further corroborated by the snapshots of the system taken in Fig.~\ref{fig:snapshots} for different system sizes, where the green sphere of radius $R^\star$ depicts the position of the ion (red symbols for bosons). Figure~\ref{fig:snapshots}(a) shows the snapshot of the system in the case when the number of bosons is smaller than the critical one, $N<N_{\mathrm{c}}$, and therefore all bosons are close to the ion forming a spatially localised MBBS. Contrarily, Fig.~\ref{fig:snapshots}(b) depicts the case with $N>N_{\mathrm{c}}$. In this scenario, the boson density around the ion is still higher than the average one and the excess bosons form a background gas.

Figure~\ref{fig:energy}(b) shows the energy dependence on $N$ following the ``polaronic'' branch where no two-body bound states are present, for three characteristic values of the atom-ion scattering lengths: $a_{\mathrm{ai}}=-0.1R^{\star}$, $a_{\mathrm{ai}}=-R^{\star}$, and the unitary case $a_{\mathrm{ai}}\to\pm\infty$. For large system sizes it is expected that the total energy can be decomposed into two contributions, the chemical potential $\mu_{\mathrm{pol}}$ of the ion and the energy of a homogenous gas, $E=\mu_{\mathrm{pol}}+Ngn/2$. For sufficiently small $|a_{\mathrm{ai}}|$, the polaron energy can be calculated variationally~\cite{ShchadilovaPRL16}
\begin{equation}
\mu_{\mathrm{pol}}=4\pi(n\,a_{\mathrm{aa}}^{3})\left(\frac{R^{\star}}{a_{\mathrm{aa}}}\right)^{2}\left(\frac{a_{\mathrm{aa}}}{a_{\mathrm{ai}}}-\frac{a_{\mathrm{aa}}}{a_{0}}\right)^{-1}E^\star\;,
\label{eq:mu:variational}
\end{equation}
where $a_{0}=\frac{32}{3\sqrt{\pi}}\sqrt{na_{\mathrm{aa}}^{3}}$ is the shift of the scattering resonance due to the bosonic ensemble. Therefore, in the regime of sufficiently weak interactions the energy is universal as it depends only on $a_{\mathrm{ai}}$ and no finite range corrections are required. 

\begin{figure}
\centering
\includegraphics[width=0.45\textwidth]{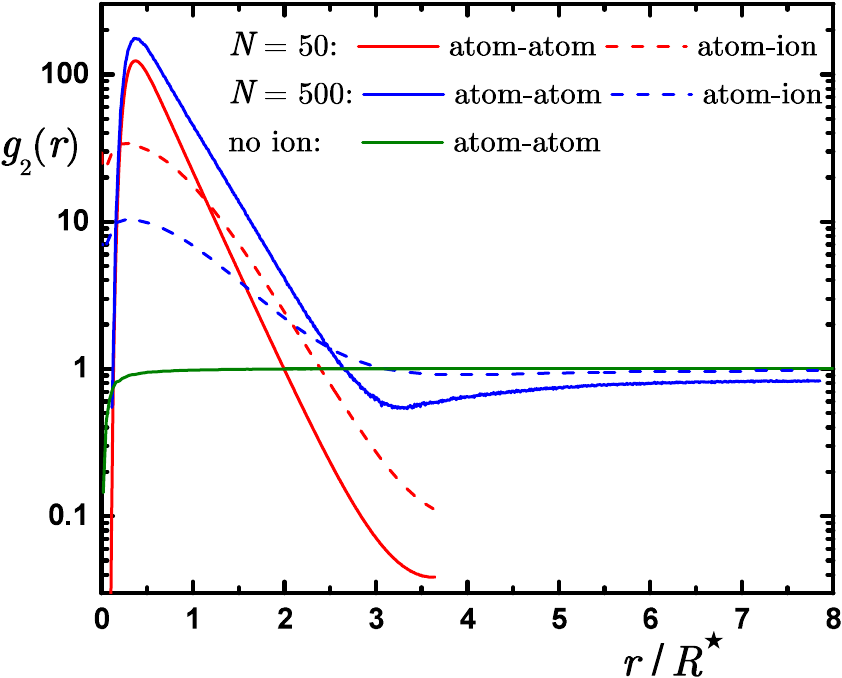}
\caption{\label{fig:corr}
{\bf Two-body correlation functions.}  
Atom-atom (solid line) and atom-ion (dashed line) two-body correlation function $g_2(r)$ obtained from variational Monte Carlo calculation for $N=500$ atoms (upper curves) and $N=50$ atom (lower curves) for $a_{\mathrm{ai}} = R^\star$; $g_2(r)$ in a weakly interacting Bose gas in absence of the ion is shown with a green line.
The critical number of atoms $N_{\mathrm{c}}$ which can be trapped by an ion is larger than the number of particles for $N=50$, which explains the pronounced maximum observed in the red lines at distances of the order of the ion potential range, $r\approx R^\star$. 
The data is shown up to the half-size of the simulation box $L/2$ and as a result the data with $N=50$ abruptly stops without yet reaching a plateau.
Instead, for $N=500$ particles, the number of atoms is larger than the critical number $N_{\mathrm{c}}$ and the half-size of the simulation box $L/2$ is larger than the size of the many-body bound state. As a result the long-range plateau observed for $N=500$ atoms signals presence of a homogeneous gas of atoms. } 
\end{figure}

It is important to notice that the Fr\"ohlich model~\cite{CasteelsJLTP11} alone is not sufficient to describe the ion quantitatively, predicting $\mu_{\mathrm{pol}}=-0.096 E^\star$ for $a_{\mathrm{ai}} = R^\star$. Instead, beyond-Fr\"ohlich perturbation theory correctly describes the polaron energy both in the weakly-interacting regime [$a_{\mathrm{ai}}=-0.1\,R^{\star}$ data in Fig.~\ref{fig:energy}(a)] and remarkably even for strongly interacting polarons ($a_{\mathrm{ai}}=-R^{\star}$). In the regime of weak interactions, the ion behaves similarly to a neutral impurity with short-range interactions, for which the VMC energy is shown with short-dashed lines. The unitary regime is reached when the atom-ion scattering length significantly exceeds the mean interparticle distance. An important feature of the ion impurity is that the energy of the many-body ground state is continuous when crossing the $a_{\mathrm{ai}}\to\infty$ point and connects the polaron and MBBS which are both stable branches. We note that directly at unitarity the prediction~(\ref{eq:mu:variational}), which is derived within Bogolyubov approximation, is beyond its validity and thus it is not shown in Fig.~\ref{fig:energy}(b). In particular, the Bogolyubov approximation becomes questionable close to unitarity, where the correlation functions shown in Fig.~\ref{fig:corr} indicate a varying local gas parameter similar to the discussion of beyond Bogolyubov corrections in Ref.~\cite{guenther2020mobile}. For the same reason, the Bogolyubov-Fr\"ohlich description of the ion polaron~\cite{CasteelsJLTP11} has to be significantly revisited.

{\it Correlation functions.}
In order to analyse the spatial structure of the many-body bound state we further turn our attention to the atom-atom and atom-ion correlation functions. Typical examples are displayed in Fig.~\ref{fig:corr}. As it can be seen, the atom-ion correlation features a pronounced peak indicating a strong bunching effect at distances where the atom-ion interaction potential is strongly attractive. Moreover, for $N < N_{\mathrm{c}}$ (see red lines) the atom-atom correlation function does not approach a constant at long distances, but instead it decays exponentially, supporting the interpretation that essentially all bosons are bound to the ion and are localized at distances of the order of a few $R^\star$. The width of $g_2^{\mathrm{ai}}$ can be used as the definition of the size of the MBBS that can be interpreted as a mesoscopic molecular ion. For $N > N_{\mathrm{c}}$ (see blue lines) the position of the peak does not change; the atom-atom correlation function, however, converges to a constant value which is slightly below unity. This demonstrates that the excess atoms are not bound to the ion and indeed form a bosonic background for the MBBS. The atom-atom correlation functions in the presence of the ion (dashed lines) also indicate the bunching behaviour close to the ion. The effect is the strongest for small systems, $N < N_{\mathrm{c}}$, where the bosons tend to stay close to each other as they are a part of the MBBS [see also Fig.~\ref{fig:snapshots}(a)]. This can be interpreted as an effective interaction within the medium induced by the impurity. As the system size is increased, $g_2^{\mathrm{aa}}(r)$ starts to approach a constant value at large distances, i.e. the whole volume is filled with the gas [see also Fig.~\ref{fig:snapshots}(b)], and the peak at short distances is correspondingly lowered. The asymptotic value $g_2^{\mathrm{aa}}(r)\to 1 - N_{\mathrm{c}}/N$ reflects a smaller effective density, as $N_{\mathrm{c}}$ atoms are bound to the ion. Eventually, in the thermodynamic limit atom-atom correlations will coincide with those of a homogeneous Bose gas without an ion (green line in Fig.~\ref{fig:corr}). 

\begin{figure}
\begin{center}
\includegraphics[width=0.45\textwidth]{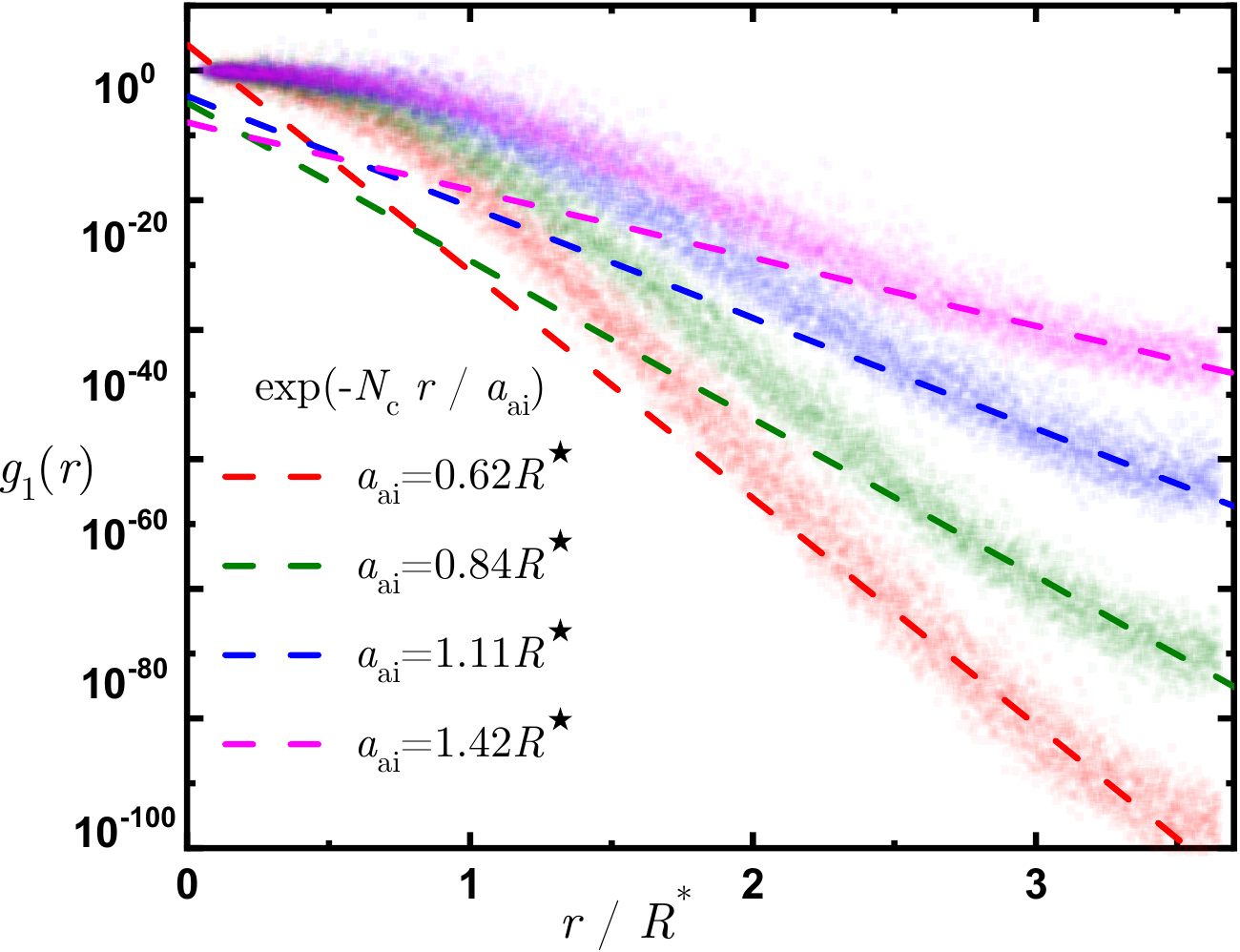}
\end{center}
\caption{
{\bf Residue function of the ion impurity. }
The symbols show the extrapolated values of the residue $g_1(r)$; the lines show the analytical approximation in terms of a decaying exponent, Eq.~(\ref{Eq:g1}). Different values of the atom-ion $s$-wave scattering length $a_{\mathrm{ai}}$ are obtained by fixing $b=0.0023R^\star$ and by changing $c$ in the ion-atom interaction potential.
Simulation is done for $N=50$ atoms. 
Extrapolation procedure based on combining variational and diffusion Monte Carlo data is used to minimize bias on the choice of the trial wave function.
}
\label{fig:g1}
\end{figure}

We have also found that for $a_{\mathrm{ai}}<0$ the two branches have very different behaviors in terms of coherence, which is quantified by the quasi-particle residue $Z=\lim_{r\to\infty}g_1(r)$
corresponding to the long-range asymptotic of the residue function $g_1(|{\bf r}-{\bf r'}|)=\langle\Psi^\dagger({\bf r})\Psi({\bf r'})\rangle$ where the field operator $\Psi^\dagger({\bf r})$ creates an ion at position ${\bf r}$ and $\langle\rangle$ denotes the ground-state average. Indeed, the residue is finite in the polaron branch and approaches unity (full coherence) in the limit of weak attraction, $a_{\mathrm{ai}}\to0^{-}$. Instead, in the MBBS branch the residue vanishes exponentially fast. Figure~\ref{fig:g1} shows typical examples of the decay of the ion residue function, $g_1(r)$, for a fixed number of particles for several choices of the atom-ion $s$-wave scattering length. We observe an exponential decay, which on the semi-logarithmic scale of Fig.~\ref{fig:g1} is seen as a linear dependence. This can be understood using a simple model. As discussed in the context of Fig.~\ref{fig:energy}(a), the energy of small clusters of bosons in the MBBS branch can be reasonably well interpreted in terms of $N$ non-interacting atoms bound to the ion. The two-body scattering solution~(\ref{Eq:scattering}) for each atom-ion pair scales as $f(r)=\exp(-r/a_{\mathrm{ai}})/r$ with $r=|{\bf R} - {\bf r}_i|$ for $r\gg R^\star$. By assuming a product over ${\bf r}_i$, $i=1,\cdots,N$ we arrive at the following approximate form for the residue function at large distances
\begin{equation}
g_1(r) \propto \exp\left(-\frac{N_{\mathrm{c}} r}{a_{\mathrm{ai}}}\right)\;.
\label{Eq:g1}
\end{equation}
For deeply bound states, $a_{\mathrm{ai}}\to 0$, all $N$ particles are bound and participate in the MBBS. In turn, for weaker interactions or larger numbers of particles, the ion is able to capture only $N_{\mathrm{c}}$ atoms. We take this effect into account by substituting $N$ by $N_{\mathrm{c}}$ in Eq.~(\ref{Eq:g1}). As it can be seen in Fig.~\ref{fig:g1}, the asymptotic expression~(\ref{Eq:g1}) captures correctly the exponential loss of coherence. 

Let us finally briefly discuss the dynamic properties of the system. In the polaronic case, the ion effective mass approaches its bare value $M^\star\approx m$ in the limit of weak attractions, whereas for stronger ones it gradually increases for the given boson-boson scattering length to the value at unitarity $M^\star \approx 6m$, which is substantially larger than the neutral impurity result $M^\star \approx 1.65m$~\cite{ArdilaPRA2015}. In the MBBS regime, $M^\star$ becomes exceedingly large. In particular, for large $E_{\mathrm{b}}$ and small $N$ we find that $M^\star \approx N_{\mathrm{c}} m$ and the total energy is given by Eq.~(\ref{eq:mu:pinned}). We could not verify whether the relation $M^\star\approx N_{\mathrm{c}} m$ holds in the thermodynamic limit due to computational limitations.

{\bf Discussion}

Our calculations are focused on the ground state properties of the system. Furthermore, we have assumed that the two-body ion-atom potential only supports one or zero bound states, while a realistic interatomic potential typically features hundreds of vibrational levels, similar to the potentials with van der Waals tails describing the interactions between neutral atoms. For the latter, however, the occupation of bound states of the interatomic potential is less likely for typical quantum gas densities, unless the system is tuned close to a Feshbach resonance, since the spatial range of the potential is on the order of a few nms, and therefore it can be well described by a pseudopotential. This is not the case for the atom-ion system, whose spatial range of the polarisation potential is tens of times larger. A natural question is then the experimental relevance and the prospects for observing the phenomena we have described.

The main process stemming from the existence of deeply bound states is the three-body recombination, which will inevitably lead to losses, as it also does for neutral Bose polarons. The timescale for such losses can be estimated with the classical trajectory result for the three-body recombination rate constant, which can be expressed as $K_3\simeq 12.52 \frac{\hbar}{m_r}\left(R^\star \right)^4 \left(E/E^\star\right)^{-3/4}$ and the decay rate given by $\gamma=K_3 n^2$~\cite{JesusPRA18,KrukowPRL16}. While for a thermal gas with density $n=10^{12}$cm$^{-3}$ and collision energies of the order of a milikelvin this gives lifetimes of the order of a second (with $\gamma\approx 2.4\,$Hz), an ion in a high density BEC is subject to much stronger losses (for $n=10^{14}$cm$^{-3}$ at $1\mu$K $\gamma\approx 140\,$kHz). This leads to submilisecond time scales, which nevertheless are sufficient to observe ion dynamics in experiments~\cite{dieterle2020transport}. For our gas parameter $\gamma\approx 600\,$Hz while the characteristic energy $E^\star/\hbar=1646\,$Hz. We further note that the quantum three-body recombination involving an ion is still not fully understood and may deviate from the classical result, e.g. it may feature minima for certain parameters (similar to loss recombination minima found for neutral atoms~\cite{BRAATEN2006259}). In particular, the dependence on the binding energy of the weakly bound state should be similar to the case of van der Waals interactions for which $K_3\propto a_{\mathrm{ai}}^4$. 

For sufficiently small loss rates, experimental detection of the signatures of the many-body bound state formation can be realized e.g. by injecting the ion into a cold gas and dragging it slowly using an external electric field, as has been done in~\cite{dieterle2020transport}. The response of the impurity and the measured time of arrival at the detector will then be mainly determined by the dramatically increased effective mass of the impurity. Moreover, one can use precise {\it in situ} imaging techniques with high spatio-temporal resolution such as the setup based on charged particle optics~\cite{Veit2020} to study the increase in the gas correlation functions due to the presence of the ion which would provide further information about MBBS formation dynamics. Radiofrequency and microwave spectroscopy developed for neutral gases can be used here as well, in particular to investigate the polaronic branch. Finally, quenching protocols in which one makes use of the ion hyperfine structure can be implemented. Taking advantage of the existence of Feshbach resonances, one can transfer an initially noninteracting ion to a superposition state with vastly different scattering lengths and perform Ramsey spectroscopy~\cite{KnapPRX12,SchmidtRPP18} to determine e.g. the quasiparticle weight. We note that most of these techniques still require some experimental progress in reaching sufficiently low temperatures to increase the interaction times and the number of partial waves involved.


In conclusion, we have investigated the ground-state properties of an ion immersed in a dilute Bose gas by means of Quantum Monte Carlo and Bogolyubov techniques. We identify three physically different regimes in the many-body system depending on the presence of the bound state in the atom-ion scattering problem: (i) polaronic branch, two-body bound state is absent; (ii) many-body bound-state (MBBS) branch, two-body bound state is present; (iii) unitarity, at the threshold of the appearance of the bound state. In the polaronic branch, many-body dressing leads to formation of a quasiparticle (ionic polaron). In the limit of weak interactions, variational methods developed for neutral atomic polarons accurately predict the energy of the system. Close to the unitarity limit the calculations unveil strong deviations from the approximate results. Finally, the MBBS branch is characterized by the formation of a large cluster (consisting of hundreds of atoms) around the ion, which in this case possesses a large effective mass, thus providing a strong analogy between the MBBS and a localized state. These quite distinct regimes should give rise to different timescales in the impurity dynamics observed in experiment, especially when combined with Feshbach resonances that allow for tuning the position of the last bound state~\cite{IdziaszekPRA09}. Our results highlight the important role of the interatomic interactions which are strongly enhanced in the proximity of the ion, driving the system away from the weakly interacting regime to a nontrivial state characterized by the interplay of long-range interaction and high local density. Apart from the atomic gases, these findings can be relevant to condensed matter systems such as electron-doped exciton gases in heterostructures of two-dimensional semiconductors, where the long-range electron-exciton interaction also has long-range character which cannot be neglected for typical experimental parameters~\cite{Fey2020}.


{\bf Methods.}

{\it Values of the parameters of the regularized potential.}
For the sake of numerical convenience we employed in our Monte Carlo simulations the regularized atom-ion potential (3) of the main text. We only considered a few specific values of the pair $(b,c)$ that are characteristic of the three regimes outlined in the diagram of Fig.~1 of the main text: weak-coupling Bose polaron (WCP), many-body bound state (MBBS), and strong-coupling Bose polaron (SCP). In table~\ref{tab:bc} we list those values in units of $R^\star$ and $E^\star$. 

\begin{table}
\begin{tabular}{|c|*{6}{c}|*{4}{c}|*{4}{c}|*{4}{c}|*{3}{c}|}
\hline
$a_{\mathrm{ai}}$ [$R^\star$] & & $b$ [$R^\star$] & & $c$ [$R^\star$] & & $E_{\mathrm{b}}\,(m=M)$ [$E^\star$] & & Regime & \\
\hline
\hline
-1.0 & & 0.0023 & & 0.1511 & &  -35  & & MBBS & \\
0.1 & & 0.0033 & & 0.1847 & &  -9.0  & & MBBS & \\
0.9 & & 0.0200 & & 0.2256 & & -1.6  & & MBBS & \\
10 & & 0.0858 & & 0.2910 & & 0.0  & & MBBS &  \\
97 & & 0.0846 & & 0.3034 & & 0.0 & &  SCP &  \\
1086 & & 0.0903 & & 0.3044 & & 0.0 & &  SCP &  \\
-1.1 & & 0.0023 & & 0.4738 & & NBS  & & SCP & \\
-0.2 & & 0.0023 & & 0.9070 & & NBS  & & WCP & \\
\hline
\end{tabular}
\caption{\label{tab:bc} 
Parameters of the regularized atom-ion interaction potential.
Parameters $b$ and $c$ of the regularized atom-ion interaction with corresponding three-dimensional $s$-wave atom-ion scattering length, $a_{\mathrm{ai}}$, at zero-energy and energy of bound state. NBS means no bound state is supported. 
The acronyms in the last column refer to: many-body bound state (MBBS); strong-coupling polaron (SCP); weak-coupling polaron (WCP). Close to unitarity the sensitivity of the $b$ and $c$ parameters is higher and therefore more digits are provided. The length and energy scales $R^\star \equiv \sqrt{m C_4/\hbar^2}$ and $E^\star\equiv \hbar^4/(m^2 C_4)$, respectively, are defined with respect to $m_{\mathrm{r}} = m/2$, i.e. equal atom and ion masses. 
}
\end{table}

{\it Trial wave functions for the Monte Carlo simulations.}

The trial wave functions are written as a pair product of Jastrow functions for both atom-atom and atom-ion correlations, featuring appropriate short and long-range asymptotic behavior [see Eq.~(4) of the main text].

The short-range part of both the atom-atom and atom-ion Jastrow function is taken from the lowest energy solution of the two-body scattering problem
\begin{align}
\label{Eq:scattering}
-\frac{\hbar^{2}}{2m_{\mathrm{r}}}\nabla^{2}\psi({\bf r})+V_{\mathrm{ai}}^r(r)\psi({\bf r})=E\psi({\bf r}),
\end{align}
where $V_{\mathrm{ai}}^r(r)$ is the corresponding interaction potential of Eq.~(3) of the main text and $m_{\mathrm{r}}$ is the reduced mass. For the atom-atom wave function we choose scattering states with energy $E=0$, whereas for the atom-ion wave function we use the exact two-body bound state with energy $E_{\mathrm{b}}$ when a bound state is present.\\

The long-range (large distance) part of the Jastrow term is taken from hydrodynamic theory. As shown by Reatto and Chester in Ref.~\cite{ReattoPR1967}, if phonons are the lowest-energy excitations in the system, the long-range behavior of the many-body wave function can be factorized as a pair-product of Jastrow functions. 

The atom-atom potential $V_{\mathrm{aa}}$ in Eq.~(2) of the main text is modelled by a repulsive soft-sphere potential: $V_{\mathrm{aa}}(r)=V_0>0$, for $r<R_{ss}$ with $R_{ss}=0.1R^\star$ and zero elsewhere. The height $V_0$ is chosen to reproduce the desired value of the $s$-wave atom-atom scattering length $a_{\mathrm{aa}}=0.02R^{\star}\ll R^\star$. We further choose the density $n(R^\star)^3 = 0.1288$, resulting in gas parameter $n(a_{\mathrm{aa}})^3 = 10^{-6}$.

{\it Mean-field estimate of the effective mass and critical number.} 
In order to formulate a self-consistent mean-field theory in the ion's frame of reference, the following wave function can be used~\cite{GROSS1962234} 
\begin{align}
\label{eq:Psi-Gross}
\!\!\!\!\Psi_{\mathrm{G}}(\mathbf{R},\mathbf{k};\mathbf{r}_1,\dots,{r}_N) \propto e^{i\mathbf{k}\cdot\mathbf{R}}\prod_{n=1}^Nf(\mathbf{r}_n - \mathbf{R})e^{i s (r_n - R)}.
\end{align}
Here, $\mathbf{k}$ is the ion momentum, $f^2$ the relative probability distribution of the position of the ion and the bosons, while $\nabla_{\mathbf{r}}s(\mathbf{r} - \mathbf{R})$ indicates the fluid velocity relative to the ion. Performing functional variation of the expectation value of the Hamiltonian~(2) of the main text, one obtains the ion effective mass~\cite{GROSS1962234,GrossJMP1963}:
\begin{align}
\label{eq:m*-Gross}
\frac{M^\star}{m_{\mathrm{r}}} = 
1 + \frac{M}{M+m_{\mathrm{r}}}
\frac{R_{\mathrm{\mu}}}{R_0} 4\pi R_{\mathrm{\mu}}^3 n.
\end{align}
Here, $R_0$ is a hard-core radius physically meaning the distance at which the atom-ion interaction starts to deviate from its long-range $\frac{C_4}{r^4}$ asymptote. Typically $R_0\sim 10\,a_0$ with $a_0\simeq 53$ pm being the Bohr radius. Furthermore, the distance $R_{\mathrm{\mu}}$ is defined as $\vert V_{\mathrm{ai}}(R_{\mathrm{\mu}})\vert = \mu$ with $\mu = g n$ the chemical potential of the bosons, from which we get
\begin{align}
R_{\mathrm{\mu}} = R^\star\left(
\frac{E^\star}{\mu}
\right)^{1/4}. 
\end{align}
 For the pair $^{87}$Rb/$^{87}$Rb$^+$ with an atomic density $n=10^{14}$ cm$^{-3}$ we obtain $R_{\mathrm{\mu}}\simeq 1.2 R^\star\simeq 6061 a_0$. Given this, the formula~(\ref{eq:m*-Gross}) predicts an effective mass $M^\star \simeq 8.4\times 10^3\, M$. Thus, the critical number of bosons bound to the ion can be estimated as: $N_{\mathrm{c}} = M^\star/m - 1$ with $M^\star$ given by Eq.~(\ref{eq:m*-Gross}) and $M=m$. 

Another estimate of $N_{\mathrm{c}}$ can be attained via rate and Gross-Pitaevskii equations~\cite{CotePRL02}. Denoting the binding energy of the two-body bound state as $E_{\mathrm{b}} = -\hbar^2 / (2 m_{\mathrm{r}}a_{\mathrm{ai}}^2)$, it can be shown that in the presence of many weakly interacting bosons $E_{\mathrm{b}}(N_{\mathrm{b}}) = E_{\mathrm{b}} [m a_{\mathrm{ai}}/(6 m_{\mathrm{r}} a_{\mathrm{aa}} N_{\mathrm{b}})]^{2/3}$. At thermal equilibrium one would expect that
\begin{align}
\label{eq:Nc-Cote}
\vert E_{\mathrm{b}}(N_{\mathrm{c}}) \vert = k_{\mathrm{B}} T 
\,\,\,\Rightarrow \,\,\,
N_{\mathrm{c}} = \frac{1}{6} \frac{m}{m_{\mathrm{r}}} \frac{a_{\mathrm{ai}}}{a_{\mathrm{aa}}}
\left(
\frac{E_{\mathrm{b}}}{k_{\mathrm{B}} T}
\right)^{3/2}.
\end{align}
For the pair $^{87}$Rb/$^{87}$Rb$^+$ with $a_{\mathrm{ai}}= R^\star$ (i.e. $E_{\mathrm{b}}\equiv E^\star$), $a_{\mathrm{aa}}= 100\, a_0$, and $T = 10\,$nK ($\ll E^\star/k_{\mathrm{B}}$), we obtain a critical number of $N_{\mathrm{c}}\simeq 372$. This number is much smaller than the previous estimate~(\ref{eq:m*-Gross}), but also does not agree with our numerical simulations. Moreover, our study predicts that $N_{\mathrm{c}}$ emerges already at zero temperature. Thus, it is not only determined by charge hopping and thermal fluctuations, but also by interaction-induced correlations. Finally, the formula~(\ref{eq:Nc-Cote}) predicts a reliance on the atom-ion scattering length as $a_{\mathrm{ai}}^4$, while our many-body analysis [see Fig. 2(a)] shows that there is almost no dependence on that length parameter. This finding highlights once more how semi-classical estimates can be quantitatively erroneous.

{\bf Author contributions.} G. A. and L. P. A. performed the Monte Carlo simulations with input from the other authors. A. N. proposed the research project with the support on scattering theory by K. J. and on polaron physics by R. S. All authors contributed equally to the analysis of the results and to the writing of the manuscript. 

{\bf Competing Interests.} The authors declare no competing interests.

{\bf Acknowledgements.} This work is supported by the Cluster of Excellence `CUI: Advanced Imaging of Matter' of the Deutsche Forschungsgemeinschaft (DFG) - EXC 2056 - project ID 390715994, 
the DFG project NE 1711/3-1, 
the DFG Excellence Cluster QuantumFrontiers, the DFG project SPP 1929 (GiRyd), the Polish National Agency for Academic Exchange (NAWA) via the Polish Returns 2019 programme, and the Spanish MINECO (FIS2017-84114-C2-1-P). The Barcelona Supercomputing Center (The Spanish National Supercomputing Center - Centro Nacional de Supercomputaci\'on) is acknowledged for the provided computational facilities (RES-FI-2019-3-0018). R. S. is supported by the DFG under Germany's Excellence Strategy -- EXC-2111 - project ID 390814868. 
G. E. A. acknowledges financial support from Secretaria d'Universitats i Recerca del Departament d'Empresa i Coneixement de la Generalitat de Catalunya, co-funded by the European Union Regional Development Fund within the ERDF Operational Program of Catalunya (project QuantumCat, ref.~001-P-001644).

{\bf Data availability.} The data that support the findings of this study are available from the corresponding author upon reasonable request.



%

\end{document}